# Radiation Risks and Mitigation in Electronic Systems

*B. Todd and S. Uznanski*
CERN, Geneva, Switzerland

**Abstract**

Electrical and electronic systems can be disturbed by radiation-induced effects. In some cases, radiation-induced effects are of a low probability and can be ignored; however, radiation effects must be considered when designing systems that have a high mean time to failure requirement, an impact on protection, and/or higher exposure to radiation. High-energy physics power systems suffer from a combination of these effects: a high mean time to failure is required, failure can impact on protection, and the proximity of systems to accelerators increases the likelihood of radiation-induced events. This paper presents the principal radiation-induced effects, and radiation environments typical to high-energy physics. It outlines a procedure for designing and validating radiation-tolerant systems using commercial off-the-shelf components. The paper ends with a worked example of radiation-tolerant power converter controls that are being developed for the Large Hadron Collider and High Luminosity-Large Hadron Collider at CERN.

## 1  Introduction to radiation-induced effects

Radiation has the potential to interfere with electronic devices and systems, creating so-called radiation-induced effects [1].

At ground level, atmospheric neutrons due to cosmic rays are a primary source of radiation. Cosmic rays are high-energy particles reaching Earth from space. These interact with Earth's atmosphere, producing a shower of particles: neutrons, protons, electrons, and many others, some of which reach the Earth's surface.

In particle accelerators, *the accelerators themselves* are sources of radiation. The Large Hadron Collider (LHC) at CERN is underground, hence it is somewhat protected against cosmic radiation. In this case, the main sources of radiation are the protons in the two circulating LHC beams. These can escape from the closed beam orbit, leaving the confines of the accelerator and be lost into local materials. There are several mechanisms for beam loss:

– intentional beam losses occur, due to the collision of beam in LHC experiments;
– direct beam losses can occur, where beam strikes collimation systems or absorbers;
– beam gas interactions can occur, where circulating beam hits residual beam gas.

A single escaping proton has the potential to create a significant shower of particles that propagate from the machine into the surrounding area. To understand the effects that this can have, one must consider the basic metal oxide semiconductor field effect transistor (MOSFET), which is the fundamental basis of electronic systems. An N-channel MOSFET is shown in Fig. 1.

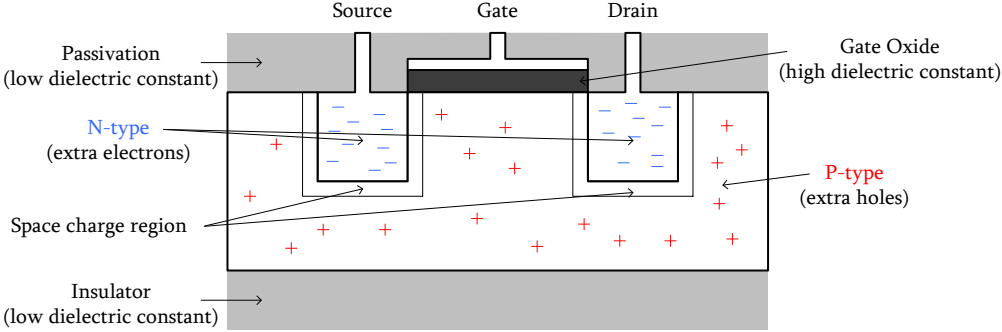

**Fig. 1:** N-MOS transistor schematic

Such electronic circuits are at risk from three principal types of radiation-induced effects: total ionizing dose (TID), single event effects (SEE) and displacement damage (DD).

## 1.1 Total ionizing dose (TID)

An incident particle can cause direct or indirect ionization of the semiconductor (Fig. 2).

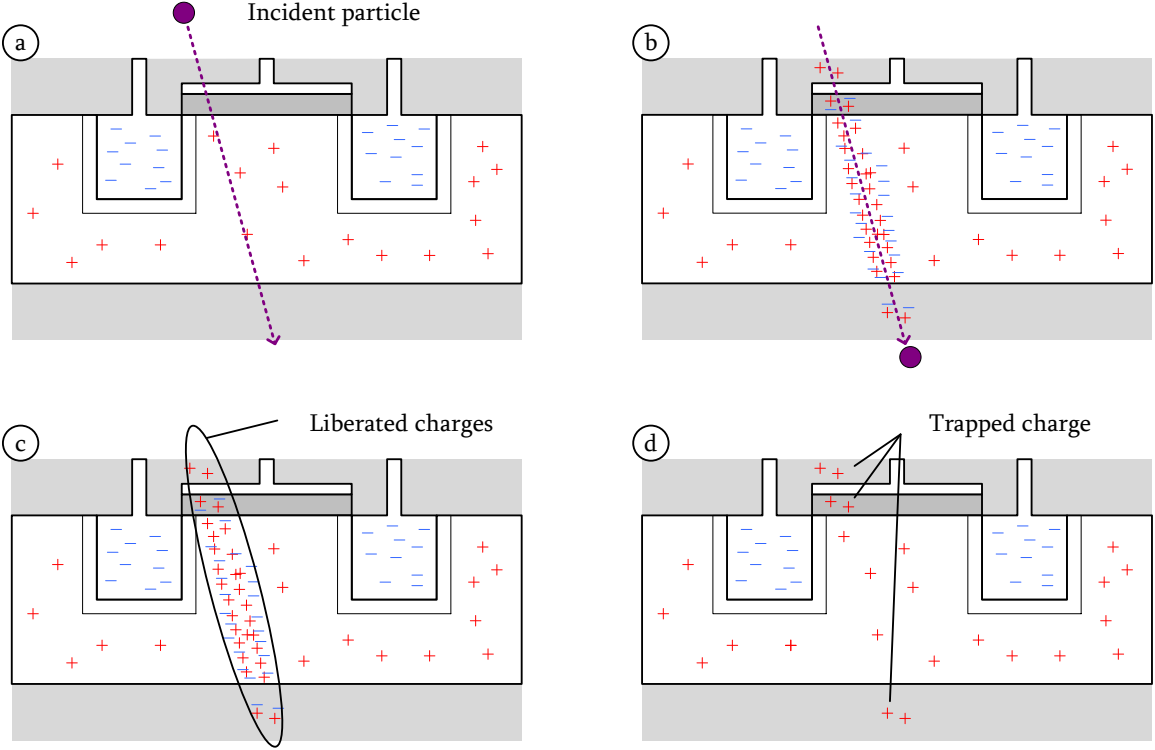

**Fig. 2:** Total ionizing dose process

Figure 2 shows an example of *direct* ionization, where the primary particle causes ionization:

(a) a charged particle passes through a semiconductor material;
(b) the charged particle interacts with the semiconductor;
(c) electron–hole pairs are generated along the path of the particle due to its energy loss;
(d) holes that are created remain trapped in the integrated circuit oxides (such as passivation, gate oxides, etc.)

*Indirect* ionization occurs when a primary particle causes nuclear reactions that then lead to ionization.

In both cases, the holes created within the oxides cause changes to device characteristics. The absorbed dose or total ionizing dose (TID) is a *cumulative* effect measured in Grays [Gy]. This increases over time causing gradual degradation of semiconductor performance.

This leads to several adverse effects. For example, consider the positive charge collected in the gate oxides: N-MOS semiconductors have a decrease in switch-on voltage, as the gate is progressively activated by the slow build-up of trapped positive charges. In the same example, P-MOS devices exhibit an increase in switch-on voltage, as the positive charges progressively inhibit the switch-on of the gate. The first observation of these effects will be changes in timing margins and increases in leakage current. Board-level power consumption can increase. Eventually, N-MOS devices will fail so that they are permanently activated, and P-MOS devices are permanently de-activated

All active components are potentially susceptible to TID.

## 1.2 Single event effects (SEE)

A single event effect (SEE) is prompt, having a certain probability (cross-section) of occurring with every particle interaction. An SEE manifests in several ways, however the root cause is generally the same (Fig. 3).

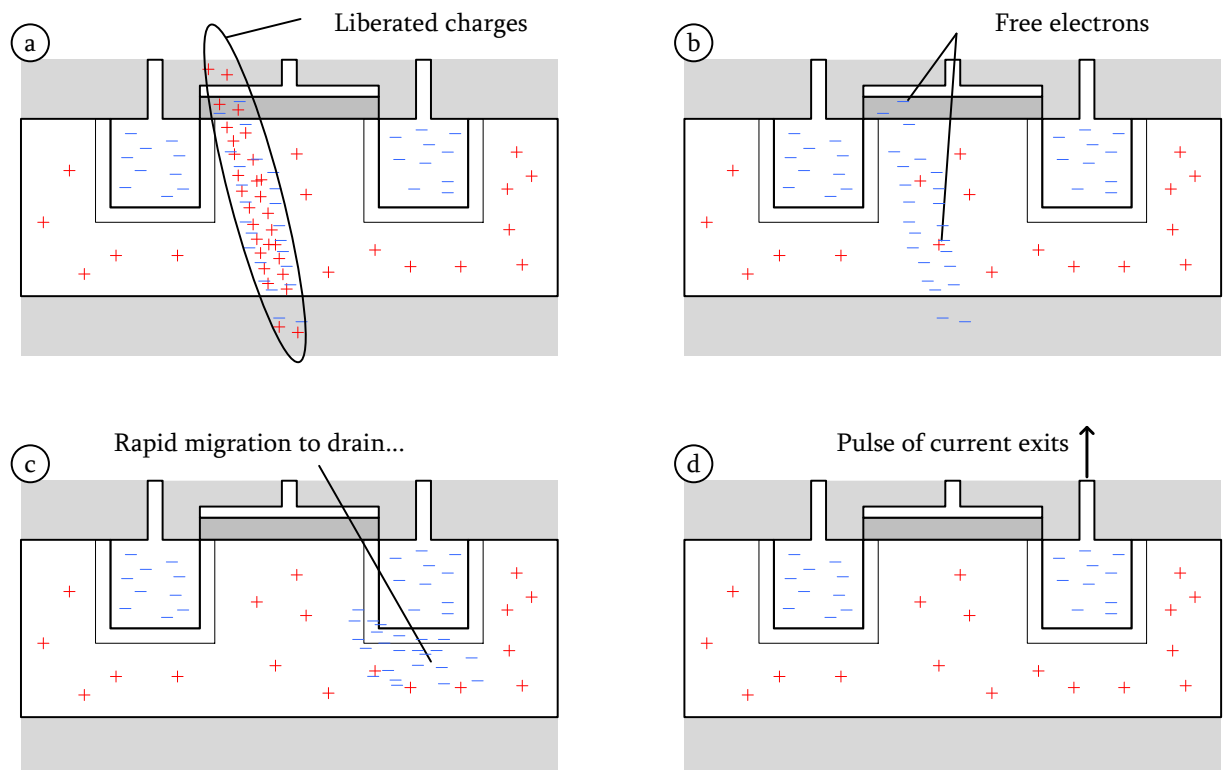

**Fig. 3:** SEE process (bias not shown, for simplicity)

Figure 3 considers an incident particle acting on a semiconductor, as described in the TID case. However, consider the behaviour of electrons, not holes:

(a) a charged particle passes through a semiconductor and interacts;

(b) electrons are generated by the particle as it traverses the material;

(c) electrons are highly mobile and flow through the MOSFET, and are collected at the reverse biased junction;

(d) these electrons create a pulse of current shortly after the particle's interaction.

In almost all cases, the root cause of the SEE is the *current pulse*. It can have a variety of impacts and outcomes on the circuit. There are two typical *soft failures*.

– Single event transient (SET) – a transient is created on a signal path. This has a variety of effects depending on the path's function. One potentially significant impact is an SET on a clock line, causing set-up and hold timing to be violated.

– Single event upset (SEU) – the initial particle interaction causes a stored bit to change in value. This erroneous value generally persists until it is re-written. The effect of an SEU on the system level depends upon the location and function of the flipped bit.

Other potentially disastrous failure modes exist, particular in complementary metal oxide semiconductor (CMOS) circuitry; consider Fig. 4.

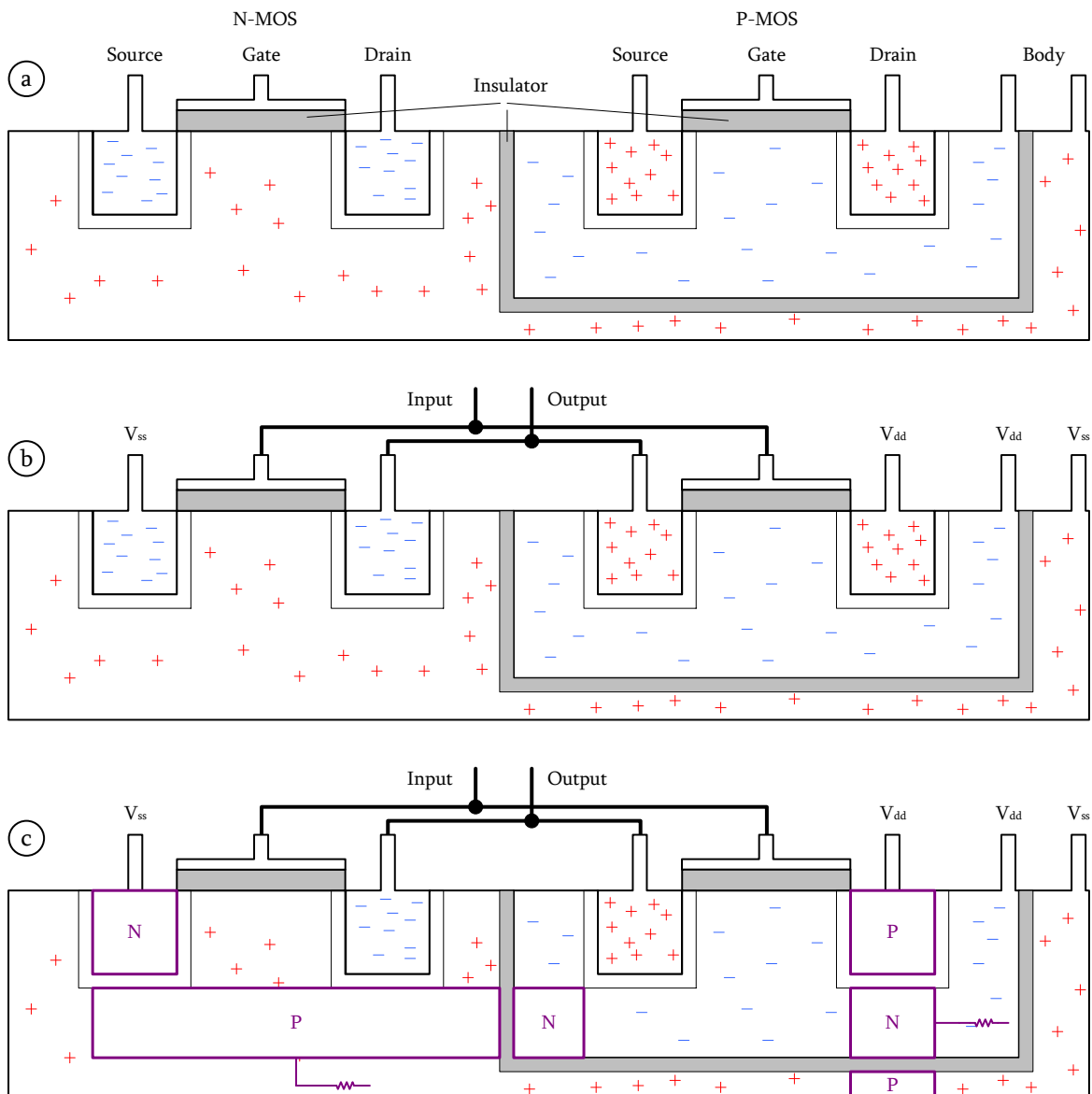

**Fig. 4:** CMOS transistor schematic

1. A CMOS structure is made from a pair of N-MOS and P-MOS devices;
2. A basic inverter configuration is shown, with the gates, sources, and drains connected appropriately;
3. A parasitic NPN, PNP structure exists in the CMOS device.

If a particle interaction deposits sufficient charge, then the parasitic structure can be activated, causing *hard failures*, such as the following.

- Single event latch-up (SEL) – a low-impedance short-circuit between power supply and ground is created; this requires power to be removed for the latch-up to be cleared. An SEL has the potential to be catastrophic if it is not mitigated, due to localized heating of the component in the high current state [2].
- SEE hard failures covers other effects such as single event burnout (SEB) and single event gate-rupture (SEGR).
- SEE are *prompt* (relative to displacement damage (DD) and TID), and each particle traversing the semiconductor has a certain probability of interaction. The particles of interest are *high energy hadrons* (HEH) or ions, and the critical parameter is the *cross-section*, which is the probability of an HEH-induced SEE.

## 1.3 Displacement damage (DD)

Displacement damage (DD) occurs when particles interact with the silicon lattice. Figure 5 shows the key process.

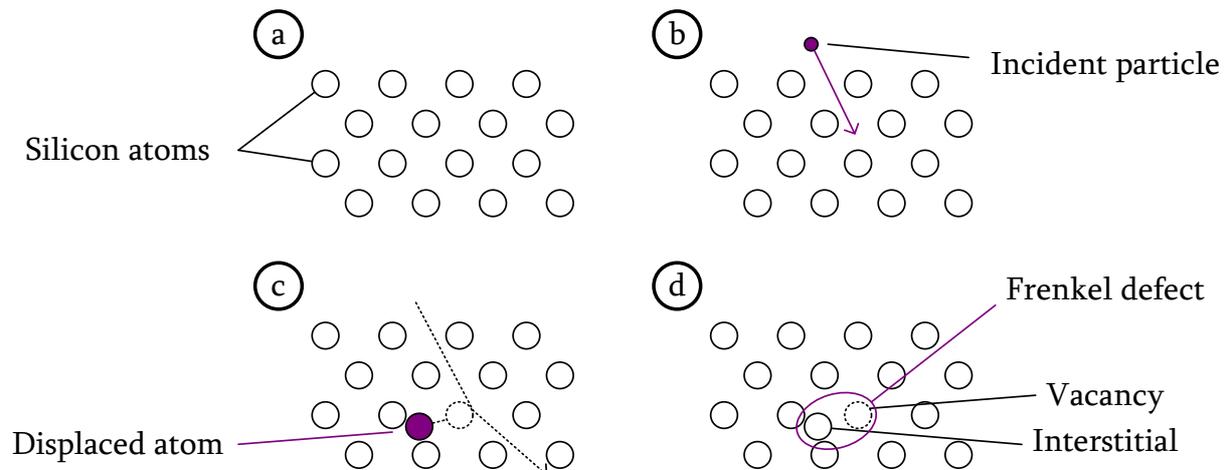

**Fig. 5:** DD process

In Fig. 5:

(a) an ideal silicon lattice has regularly spaced atoms;
(b) incident particles traverse the lattice;
(c) there is a probability that the particle strikes and dislodges an atom;
(d) this interaction can create *Frenkel defects* consisting of a *vacancy* and an *interstitial defect*.

Damage to the lattice is proportional to the integrated flux of particles that have passed through the atomic structure, therefore DD gradually increases over time. The effects of DD generally reduce device gain, which eventually leads to complete failure of the device. DD mainly affects bipolar technologies; CMOS technologies generally do not suffer from DD effects.

DD is a *cumulative* effect and is measured as the total number of 1 MeV equivalent neutrons that have passed through a given surface area [1 MeV eq. $n$/cm$^2$]. The term non-ionizing energy loss (NEIL) is also used.

## 2 Design for radiation

Dealing with radiation effects on electronic systems is a fundamental part of electronics design; at the same time, it is a tremendously complex and expensive requirement. Engineers should make every effort to first optimize the system requirements before embarking on the development of radiation-tolerant designs. Only after exhausting all alternatives should the design of a radiation-exposed system be undertaken. The main considerations to address are given below.

- Functional – is the system needed at all? Engineers should consider whether it is possible to design-out the system's function at a higher level. Are there alternative means to achieve the same goal without having a dedicated system? What are the consequences of not having this system?

- Environmental – does the system need to be here? Before undertaking a system design, exhaust all options for moving systems away from radiation, or shielding equipment from radiation. Move every element that can be moved, in order to minimize the amount of radiation-exposed elements as possible.

- Dependability – can the permitted failure rate of the system be increased? Dependability requirements should be established at an early phase. The system should have a reasonable failure rate. In addition, the permitted failure modes should be outlined. Protection-related systems may have a strict *unsafe* failure rate specification. Systems with such a requirement add further complexity to an already complex process. Engineers should make every effort to increase the permitted failure rate, and to exclude undesired failure modes.

### 2.1 Radiation hardened vs. radiation tolerant

Once it is decided that a radiation-exposed system must be designed, a primary factor is to determine whether a radiation-*hardened* or radiation-*tolerant* design is needed.

The principal consideration is the environment. Figure 6 shows typical values of radiation fluence and dose for sea, air, and space applications, as well as the approaches needed. The delimitation of the areas is not a fixed value, and can be debated.

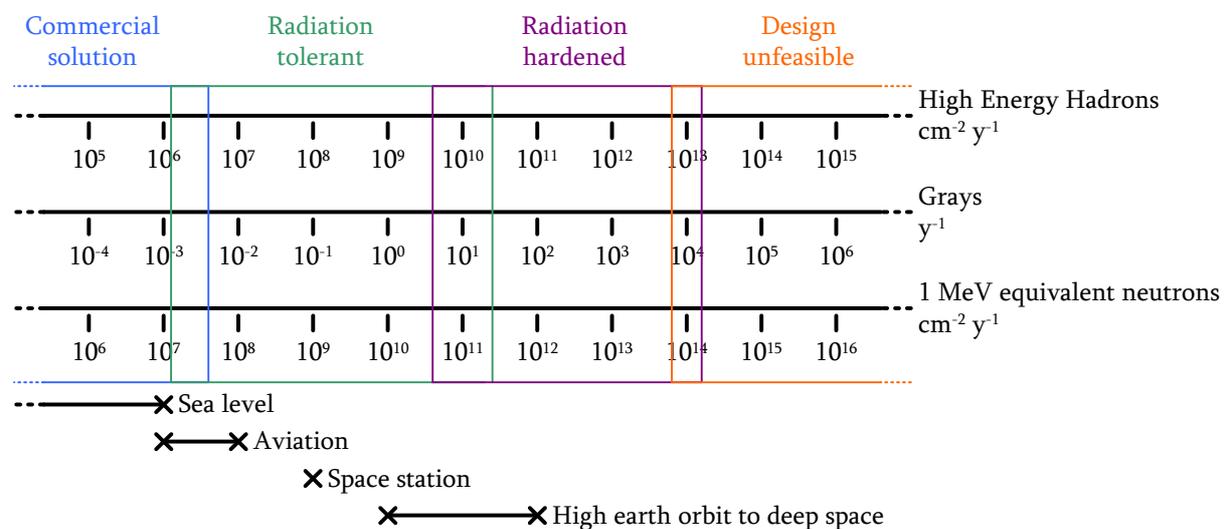

**Fig. 6:** Radiation exposure levels vs. engineering approach (CERN defined)

*Radiation-hardened* systems (as defined at CERN) exhibit no adverse effects from radiation, as they are immune to radiation-induced problems, in the environment specified, up to defined limits. These systems can make use of radiation-hardened technologies (so-called radiation-hardening-by-process) or mitigation of radiation effects at the design level (radiation-hardening-by-design). A complete, certified, radiation-hardened hardware solution can be designed. One of the key drawbacks of this approach is cost. For example, a radiation-hardened programmable logic device is 100–1000 times more expensive than a commercial device [3], and a custom-made integrated circuit can have extremely high set-up costs in the order of millions of Swiss Francs [4].

*Radiation-tolerant* systems (as defined at CERN) are those that can be made to function in radiation environments despite being susceptible to radiation. This is primarily done by mitigating the consequences of the effects. This paper explains the process of designing *radiation-tolerant* systems based on commercial off-the-shelf (COTS) components.

## 2.2 Radiation-tolerant design flow

A design flow has been created to address project risks specific to the development of radiation-tolerant COTS electronics. The design flow begins after the *requirements capture* and *engineering specification* phases. The process flows of electronics system and radiation-tolerant system development are shown in Fig. 7.

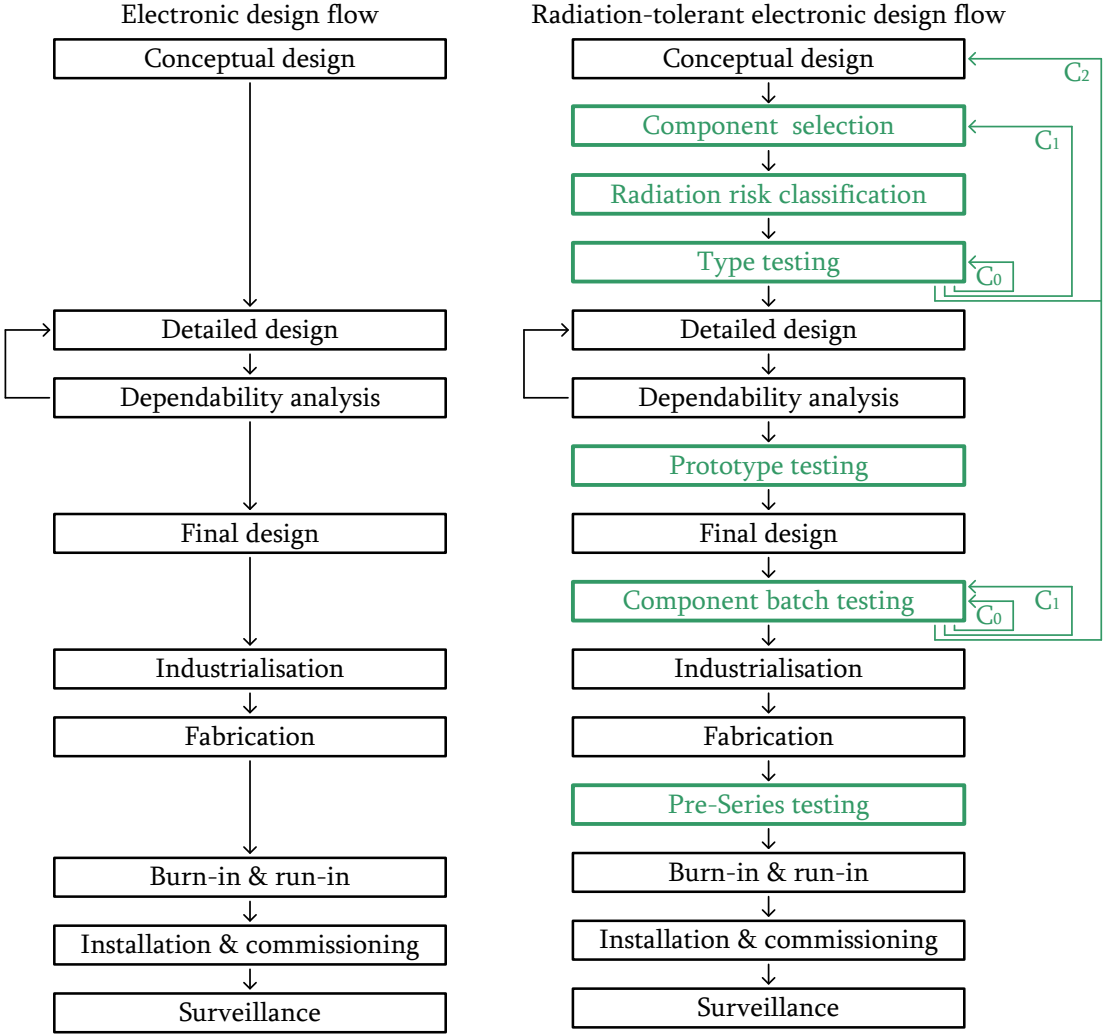

**Fig. 7:** Electronics design flow and radiation tolerance electronics design flow [5, 6]

The arrows in Fig. 7 show the impact of an electronic component that is not satisfactory. The arrows are marked $C_0$, $C_1$ and $C_2$ as a function of criticality of the unsatisfactory component in the design flow. A complete classification of components, their criticalities and different steps are described in the following paragraphs of this section. This flow can be directly compared with the Radiation Hardness Assurance of components ECSS-Q-ST-60-15C [7].

- Conceptual design – a basic design for the system is selected, with solutions proposed for the implementation of each required function from the *engineering specification*.

- Component selection – the *conceptual design* is broken into sub-systems, and a component inventory is established, with components organized by functional requirements. The component selection and creation of the bill-of-materials (BoM) are made in close collaboration with the design team. The defined functionality of the system needs to be balanced against the feasible radiation-tolerant solutions. This is an iterative process, which continues as radiation tests advance. Some components may need to be changed as results appear from testing. Radiation testing is a slow and complex process, thus minimization of component count and variation is a key requirement.

- Radiation risk classification – each component to be used is classified into one of three risk classes ($C_{0-2}$) based on a risk matrix, cross-referencing the likelihood and impact of radiation-induced failure. Three main criteria are used for the component classification:

  1. the known susceptibility of the type of component to radiation;
  2. the function of this component in the design, i.e. the impact of its failure on system reliability;
  3. the availability of alternatives for the component.

A summary is given below.

- $C_0$ components are those that are known to be resistant to radiation, or are easily replaced if found to be weak. In general the design of the system is not influenced by these $C_0$ parts. Examples: resistors, capacitors, diodes, transistors, etc.

- $C_1$ components are those that are potentially susceptible to radiation, in less critical parts of the system. Substitution of parts or mitigation of issues is possible with a redesign. Examples: regulators, references, drivers, level translators, memory, etc.

- $C_2$ components are those that are potentially susceptible to radiation, being found in more critical parts of the system. If these parts do not perform well then the basic design is compromised, and a significant rework of the *conceptual design* is needed. Substitution of parts or mitigation of issues due to $C_2$ parts is generally difficult. Examples: analog-to-digital converters (ADCs), programmable logic, mixed circuits, etc.

- Type testing – the selected components are subject to radiation testing to identify cross-sections and to predict component lifetimes. Various sources of radiation test data can be consulted from other institutes and organizations such as the ESA [8], NASA [9, 10] or in the IEEE NSREC data workshop proceedings. In some cases, test setup, test procedure and test parameters such as bias conditions or temperature may be similar to the project application and the test results can be directly used without a dedicated radiation characterization test. It should be noted that some of the ESA/NASA tests are performed on high-reliability electronic components, which can have different radiation tolerance to regular COTS components, even when considering the same reference numbers from the same manufacturer.

In the absence of reliable radiation results for a given component, a radiation campaign is prepared. Each individual component is tested by applying procedures that are briefly described in Ref. [11]. The higher the class of the component, the more detailed is the analysis of its response to radiation.

Several reference test standards exist for the planning and execution of the radiation tests. These are: for SEE, JEDEC Test Standard 57 [12] or ESA/SCC 25100 [13]; for TID, MIL-STD 1019.8 [14], ASTM F1892 [15] or ESA/SCC 22900 [16]. An extensive review of TID radiation hardness assurance standards can be found in Ref. [17]. The level of testing required depends upon the class of the component (Table 1).

Table 1: Radiation characterization test methodology

| Class | Mixed field | Proton | Heavy ion |
|---|---|---|---|
| $C_0$ | Mandatory | | |
| $C_1$ | Optional | Mandatory | |
| $C_2$ | Optional | Mandatory | Mandatory |

At CERN, $C_0$ components are tested in a mixed-field radiation environment equivalent to LHC tunnel conditions, thus giving a direct indication of the device's performance in the final application.

$C_1$ components are irradiated with mono-energetic protons to measure susceptibility to SEE and TID; mixed-field tests are optional proton tests that allow assessment of SEE and TID response of a component in a much shorter time. Two or more candidate parts for each $C_1$ position should pass *type testing*. If this cannot be ensured, it is possible that a return to *component selection* is needed to find more candidate parts.

Type-testing failure of $C_2$ parts is critical for the project. These are tested in exactly the same way as the $C_1$ components, with the addition of heavy-ion testing to assess SEL cross-section and the respective risk in the LHC radiation environment. At least one $C_2$ part must be found that meets system requirements. If this is not the case the *conceptual design* must be revised.

– Detailed design – the remainder of the design is established, using components that performed well in *type testing*.

– Dependability analysis – the reliability of the system hardware is predicted, using cross-sections, lifetimes and electrical characteristics. Traditional reliability engineering techniques can be used to meet system requirements.

– Prototype testing – a small number of prototypes are constructed, following the concept emerging from the dependability analysis. These are used to validate design choices, and to ensure correct integration between connecting systems and controls.

– Final design – the final design is established, which *on paper* meets all of the functional and dependability requirements of the system.

– Component batch testing – after prototype validation and definition of the final implementation, the procurement of large quantities of components begins. Each has to be qualified to confirm the results of the type testing and to assure the conformity of the component radiation response within the lot. COTS components form the main part of those used in this framework; ideally all production components should be procured from a single fabrication lot to decrease the component-to-component variability. In many cases, it is impossible to get such information concerning the number of silicon wafers from which the components were made, their lot date code, or even the lot origin. This makes lot acceptance tests complex and challenging. ESA specifications require a minimum of 11 samples to be selected for TID characterization: ten for irradiation and one reference part. Similarly, the CERN lot acceptance test strategy requires a minimum of ten irradiated samples and one reference for each component lot. The number of samples to be tested for SEE characterization is much smaller, typically three [13].

If a $C_0$ component does not pass testing, another equivalent component will be purchased and lot acceptance tests will be performed. If a $C_1$ component does not pass the lot acceptance tests, its equivalent will be chosen from a list of preferred replacements for $C_1$ components prepared in advance during preparation of the BoM. As shown in Fig. 2, $C_2$ components are highly critical for the design, and in the case of lot non-conformity the project's conceptual design will have to be revised. For all $C_2$ components it is of the utmost priority to decrease the probability of lot problems.

– Industrialization and fabrication – the number of required units is produced, after industrialization, ensuring the best quality is achieved for the product being designed.
– Pre-series testing – production is usually split into pre-series and series production. A small number of parts are delivered first, which are manufactured using the industrial assembly line. Accelerated lifetime testing and dedicated irradiation testing can be carried out to ensure that industrialized fabrication yields parts that meet predictions.
– Burn-in → surveillance – the usual steps in the delivery of an engineering product.

## 3 Radiation testing

Radiation testing is a crucial aspect of the development of radiation-tolerant electronic systems. The three key radiation effects need to be characterized and quantified for each of the parts being used.

To determine the effects of *displacement damage*, parts are placed in the vicinity of radiation sources, typically neutrons. The parts do not need to be powered for displacement damage to occur; after irradiation parts are characterized on electrical test benches. Dedicated irradiation areas can be found in the proximity to fission reactors, such as PROSPERO [18].

SEE cross-sections and TID limits are determined by controlled exposure to radiation. This is done by building dedicated test equipment that exercises components whilst they are being irradiated. Tests and observations are carried out, looking for the malfunction characteristics of SEEs, and the end-of-life due to TID. Special care has to be taken to subject parts to a representative radiation field. The irradiation spectra have to give meaningful results considering the environment in which the system is to be used.

An example of a representative SEE and TID test area was the CERN Neutrinos to Gran Sasso (CNGS) gallery, which was operated until 2012. Proton beams from the Super Proton Synchrotron (SPS) were extracted onto a target, creating a neutrino beam. A radiation field is created by the interaction of the proton beam with the target, so components could be tested in the immediate vicinity (Fig. 8).

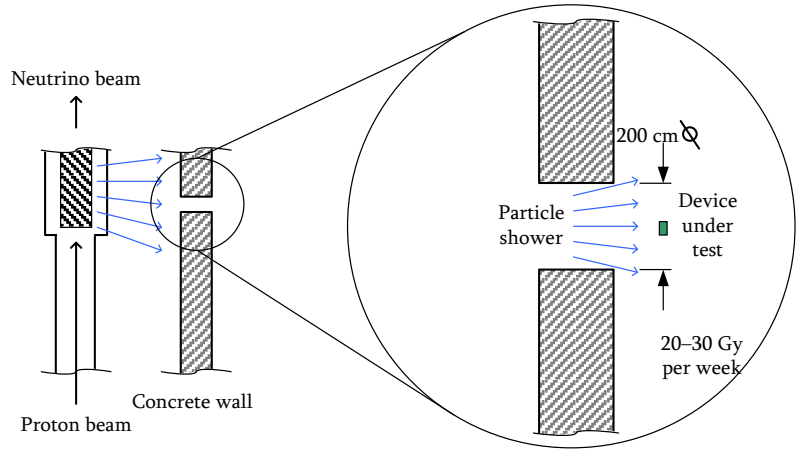

**Fig. 8:** Plan view of the CNGS installation [19]

A typical test apparatus placed in this location includes devices under test, which are surveyed and characterized. The conclusions from such tests are the device *failure modes*, and *failure mode ratios*. This is an exhaustive list of errors that have been observed, grouped by failure mode. The ratios of each type of error are also determined. In addition the TID limit is determined, giving the effective device lifetime in this particular radiation field. A drawback of these tests is limited beam availability and long irradiation time due to the low fluences that can be obtained [18]. At CERN a new facility, called CHARM, is under construction to overcome these limitations [20, 21]. CHARM is intended to provide:

- mixed-field particle spectra representative of the LHC tunnel, space, and/or ground level;
- a large testing volume to allow the testing of several cubic metres of equipment;
- high beam availability and intensity higher than an operational environment as in the LHC tunnel.

## 4 Applying the design flow to power converter controls

The design flow outlined in the previous section is being applied to the development of the power converter controls for the LHC at CERN. The LHC has several thousand normal and superconducting magnets with associated power converters. A typical example circuit is the 13 kA dipole circuit as shown in Fig. 9.

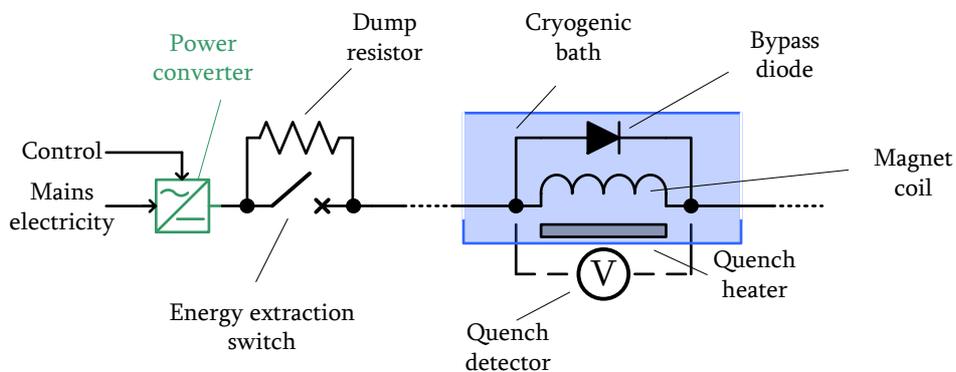

**Fig. 9:** Superconducting electrical circuit, with power converter

These circuits have a high stored energy, and use a system of interlocks and protection to guard against the uncontrolled release of magnet and powering energy. The power converter converts mains energy to the required magnetic field for beam operations (Table 2). The set point and operation of the power converter is managed by a power converter *controller*.

**Table 2: Types of power converter in the LHC**

| Principal application [magnet] | Voltage [V] | Current [A] | Quantity |
|---|---|---|---|
| Main dipoles | 13 000 | 190 | 8 |
| Main quadrupoles | 13 000 | 18 | 16 |
| Quadrupole circuits | 4 000–6 000–8 000 | 8 | 188 |
| Warm circuits | 1 000 | 450–950 | 16 |
| Sextupole circuits | 600 | 40 | 37 |
| Octupole circuits | 600 | 10 | 400 |
| Orbit correctors | 120 | 10 | 298 |
| Orbit correctors | 60 | 8 | 752 |

This gives a total of *over 1700* power converters used in the LHC. These are located in one of five different areas (Fig. 10), each with a different risk classification for radiation:

1. at the surface;
2. in parallel galleries;
3. in perpendicular galleries;
4. in alcoves;
5. the LHC accelerator tunnel.

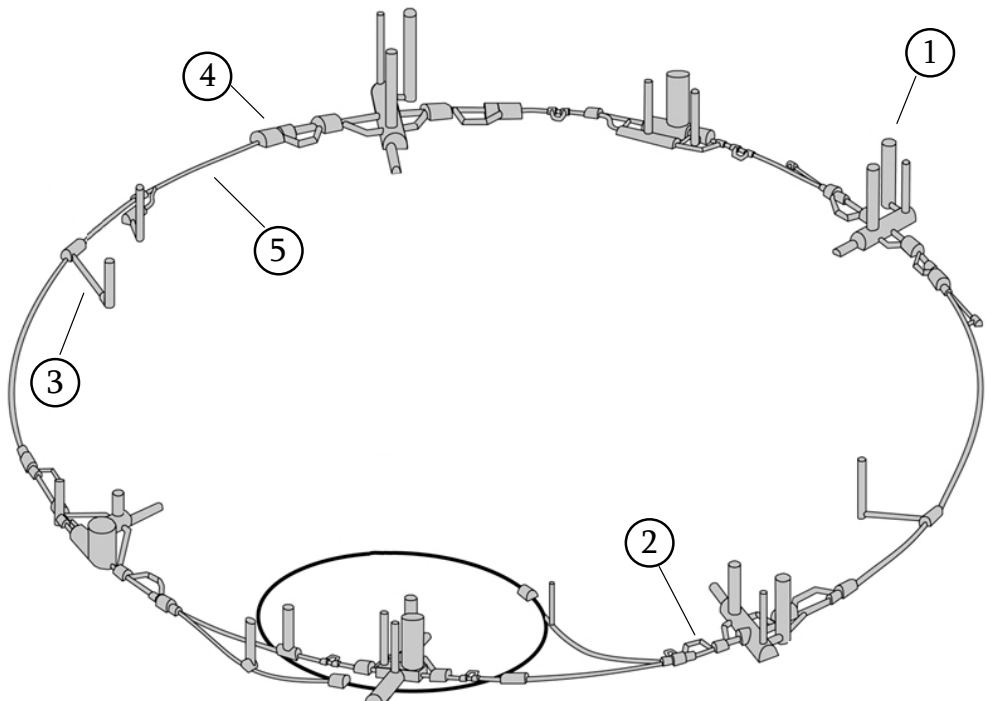

**Fig. 10:** Civil works with principal converter locations [22]

Perpendicular galleries, alcoves, and the accelerator tunnel each present a moderate to high level of risk concerning exposure to radiation. Of the power converter quantities shown before, *over 1000* are located in such areas (Table 3).

Table 3: **Power converters found in moderate to high radiation risk areas [23]**

| Principal application [magnet] | Voltage [V] | Current [A] | Perpendicular | Alcove | Tunnel |
|---|---|---|---|---|---|
| Quadrupole circuits | 4000–6000–8000 | 8 | 6 | 60 | - |
| Sextupole circuits | 600 | 40 | - | 12 | - |
| Octupole circuits | 600 | 10 | 24 | 104 | - |
| Orbit correctors | 120 | 10 | 15 | 92 | - |
| Orbit correctors | 60 | 8 | - | - | 752 |

Each power converter can be broken down into three distinct sections (Fig. 11).

1. a function generator controller (FGC) electronic module;
2. a voltage source (VS), consisting of power electronics and power circuits converting mains power to the current and voltage requested by the FGC;
3. current transformers (DCCT) converting the electrical output current of the converter into a digital value that is read back by the FGC.

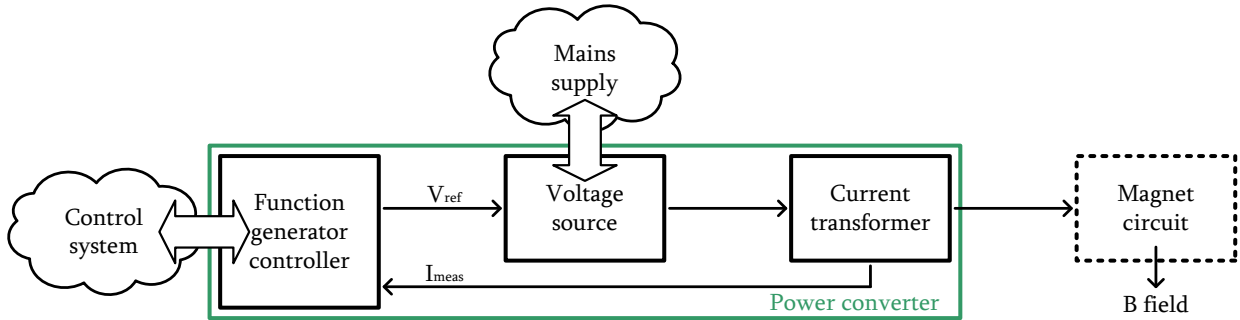

**Fig. 11:** Key components of a power converter

The design flow has been applied to the design and realization of the FGC that is to be used in the power converters located in moderate to high radiation risk areas, as shown in Table 3.

## 4.1 Function generator controller

The FGC is a purpose-built electronic module having several functions (Fig. 12). Most notably [24] it:

– implements closed-loop regulation of the magnet current, reading the measured current $I_{meas}$ to establish the reference voltage $V_{ref}$ needed for the field;

– controls the converter, by issuing digital commands such as ON, OFF, and RESET;

– implements low-level interlock logic as part of the interlock loops between the VS, quench detection, and powering interlock systems [25];

– allows remote control and surveillance of the converter and associated subsystems.

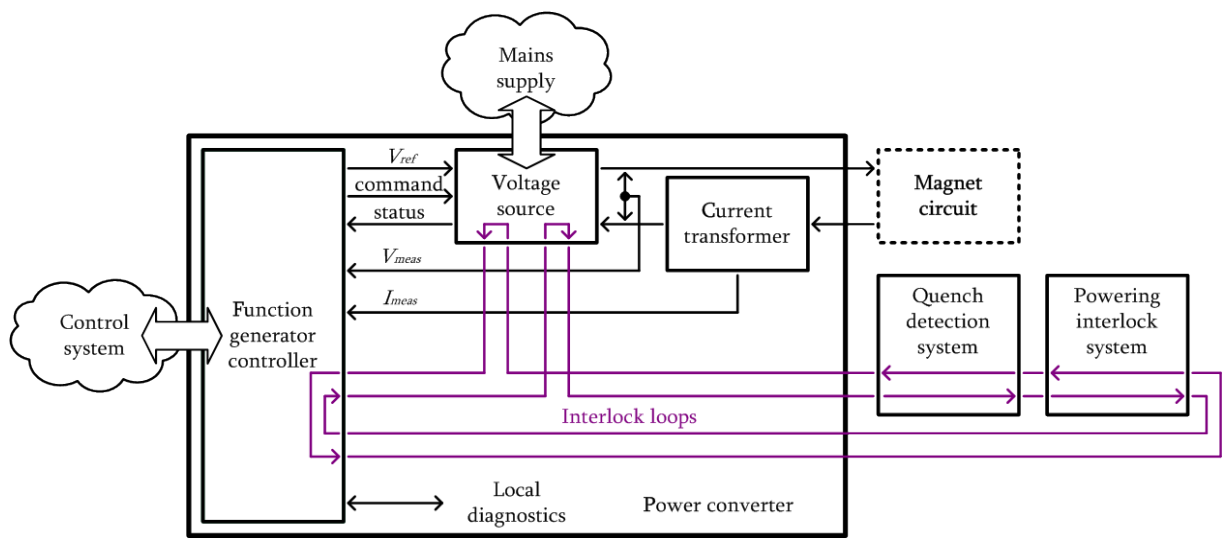

**Fig. 12:** Basic connectivity of an FGC

The control system of the LHC power converter controls is based on the WorldFIP fieldbus (Fig. 13), using a gateway computer with bus master to send and receive real-time commands to FGC slaves.

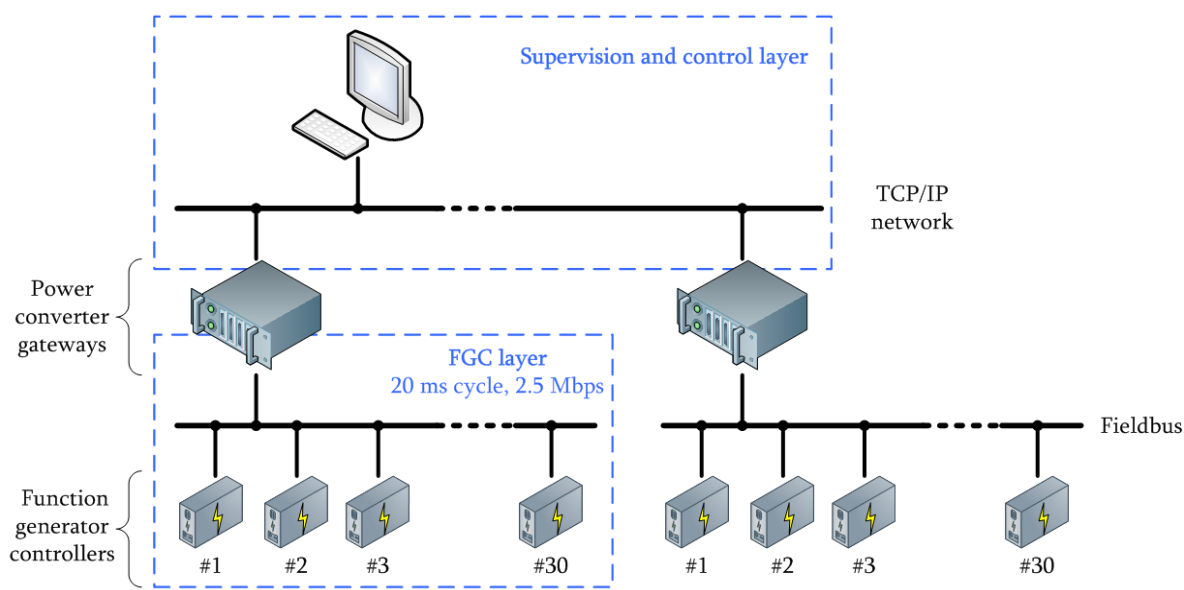

**Fig. 13:** Fieldbus control principle

The use of the fieldbus provides a low-bandwidth communications solution for the FGC. During each cycle only around 32 bytes can be transmitted to, and 128 bytes received from, each FGC [26].

## 4.2 Radiation-tolerant function generator controller

From the operational point of view, power converters are to behave in the same way regardless of whether an FGC or an FGClite controller is used. To optimize costs, the existing fieldbus infrastructure will be re-used and FGClites will be plug-compatible with FGC2. This makes significant savings but means that fundamental changes to the FGC philosophy are not possible. Effort has been put into the optimization of software, programmable logic, and hardware partitions to minimize the complexity of the FGClites, whilst meeting system-level requirements.

In FGC2, the current reference as a function of time is stored locally in each FGC. A function table and regulation circuit use circuit settings specific to the magnet circuit being powered to drive the voltage reference point (Fig. 14).

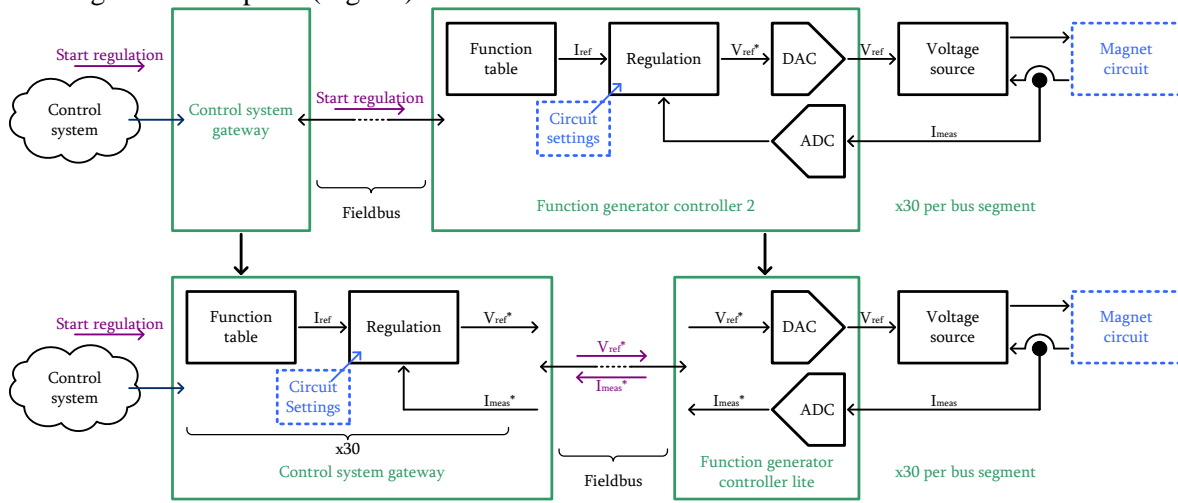

**Fig. 14:** Closed loop control architecture changes [26]

The gateway sends simple commands such as *start regulation* to each FGC. The most significant hardware change for the FGClites is the relocation of the digital signal processor (DSP) into the gateway, with the FGClite acting as a remote input/output module.

The gateway complexity is significantly increased as it is required to implement the regulation calculations for all FGClites connected on the same fieldbus segment. This increases the latency of the regulation loop due to the transmission of information back and forth on the fieldbus, which requires the regulation algorithms to be adjusted. Additionally, FGC2 was capable of working independently of the fieldbus for short periods, whereas the FGClites will be completely dependent on the fieldbus for correct operation.

### 4.3 Software and programmable logic partitioning

FGC2 depends on both software and programmable logic to achieve its functional requirements. Embedded software is used both for closed-loop signal processing and converter supervision. In the FGC2, eleven programmable logic devices are used for sub-functions such as timer circuits, access to coefficients, and digital multiplexing, amongst others (Fig. 15).

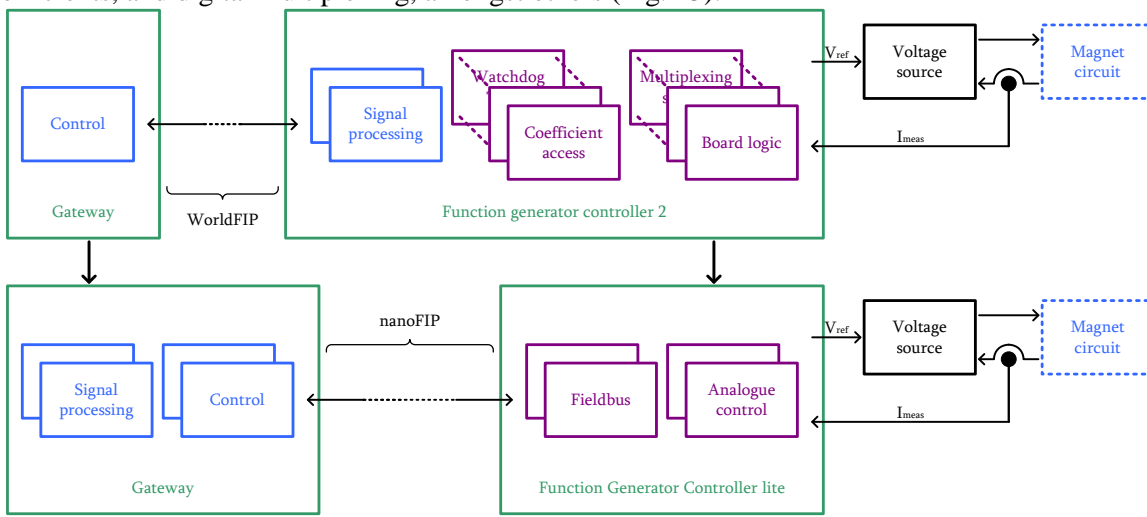

**Fig. 15:** Architecture changes

FGClites are to have no locally executed software as the signal processing functions are to be moved to the gateway. The remaining supervision requirements, as well as FGC2 functions implemented in programmable logic, are to be implemented in three flash-based field programmable gate arrays (FPGAs) having functionality described in VHSIC hardware description language (VHDL).

The obsolete WorldFIP chipset is also to be replaced by CERN nanoFIP, which also uses a flash FPGA. NanoFIP exploits the WorldFIP electrical standard but a simple transmission protocol between gateway and power converter [27].

## 4.4 Predicted reliability and lifetime

Failures of the power converter hardware can be split into two categories: basic failures corresponding to the bathtub curve and those related to radiation damage.

## 4.5 Basic failures

The first source of failure is expected to follow the typical hazard function for the failure of non-complex electronic systems, the so-called bathtub curve, made up of three sections.

- *Early-life* failures are caused by latent defects and are avoided by processes such as stress screening and running-in.

  The *useful-life* failure rate is one of the biggest concerns for the success of the FGClite project in meeting its reliability goal. Failures of this nature can occur at any moment in time and are not correlated. This is to be minimized by following design practices promoting reliability, such as over-specification and redundancy. The base failure rate of the FGClites will be determined using a combination of past experience and military handbooks.

- *Wear-out* failures are due to the gradual wear-and-tear of electronic systems in use. In the FGClites these are to be minimized by following the most appropriate maintenance plan, either preventive or reliability centred maintenance (RCM).

## 4.6 Radiation-induced failures

The second source of failure is that related to radiation. Radiation-induced damage manifests itself in two manners: cumulative and prompt. Cumulative or total dose effects reduce the effective system lifetime by advancing the wear-out phase, and SEEs increase the random-in-time failure rate of the system across its whole lifetime.

First prompt effects, such as SEU or SEL, can cause the system to malfunction, having the effect of increasing the random-in-time failure rate of the FGClites. The predicted number of failures per year can be seen as a function of the fluence of particles in the areas in which the FGC is installed, and the cross-section of each FGClite. Figure 16 shows the characteristics for a subset of converters with an LS1–LS2 estimated fluence of $9 \times 10^9$ high energy hadrons (HEH) per square centimetre per year in tunnel installations [28].

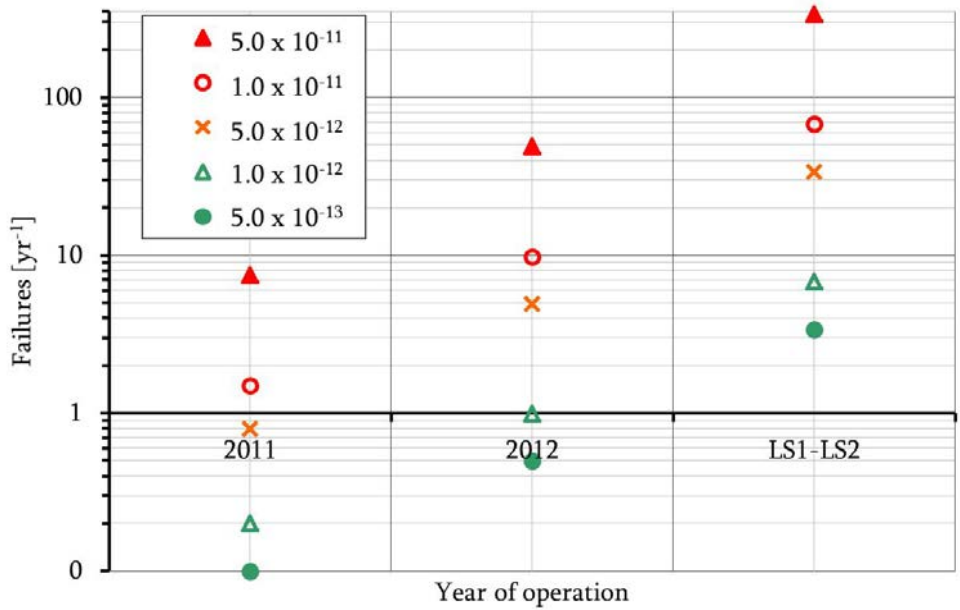

**Fig. 16:** Cross-section and predicted failure rate for LHC operational phases

### 4.7 Requirements

There are two key requirements: *lifetime*, and *reliability*.

- Lifetime: the FGClites must be designed to outlast the LHC. Current planning extends into the 2030s, so an FGClite installed in 2015 would need >25 years' lifetime. In addition to the electrical requirement, radiation dictates that every component in the FGClites must remain within specification after absorbing around 200 Gy.
- Reliability: power converters have a direct influence on the availability of the LHC machine. the LHC is expected to run for 200 days every year, with two ten-hour fills, and two recovery periods of two hours every day [29]. This gives 400 LHC missions per year. No more than 10% of these should be aborted due to the failure of power converters; this means that each power converter must have an MTBF in excess of 400 000 hours.

This can be split between electrical and radiation-induced failures. Electrical MTBF is therefore required to be similar to the existing controllers, with a maximum of 10 radiation-induced failures per year of operation for all installed systems. This means the SEE cross-section of the FGClites is required to be equal to $3 \times 10^{-12}$ cm$^2$ or lower.

Combining these requirements means that all FGClites in operation are expected to fail less than 10 times per year due to radiation-induced errors, and less than 30 times due to electrical effects, meeting the combined requirement of less than 40 failures per year.

### 4.8 FGClite project risks

FGClites are required to be installed in the LHC at the end of 2015. The most significant risk to the successful completion of the project concerns class $C_2$ components: optimization of component selection has yielded only three $C_2$ parts:

- the ADC used to determine $I_{\text{meas}}$;
- the mixed analog–digital IC used for the fieldbus interface;
- the flash-FPGA used throughout the design.

Early efforts focused on the *type testing* and *component batch testing* of these parts to determine their suitability for the FGClites.

The quality of statistics is critical for reliability calculations, as they drive both the mitigation techniques and overall FGClite reliability. Of particular importance is the predicted HEH fluence in the LHC tunnel. In this context there is a risk of over-engineering the FGClites by taking an excessively pessimistic view: layers of redundancy and power-cycling options could be in excess of the project needs, reducing overall reliability.

The shift away from software towards programmable logic has many implications, ranging from the skills required from the project team, to quality assurance of the FGClites. Reliability calculations explained in this paper assume a non-complex system, free from systematic faults. The programmable logic engineering must be of the highest quality, matching that used elsewhere at CERN, following guidelines for dependable VHDL design that have been developed in the course of other systems' developments at CERN [26].

## 5  Conclusions

This paper has explained the principal effects of radiation on electronic components, and has described the principal steps needed to design a radiation-tolerant system. The paper included a worked example showing how radiation-tolerant power converter controls are being developed.

**Acknowledgements**

This paper has been written in collaboration between the power conversion group of the CERN technology department (TE/EPC) and the CERN Radiation Effects to Electronics working group (R2E). The authors wish to express their gratitude to the numerous individuals who have helped to define this procedure, and who are active in the execution of the FGClite project.

Elements of this paper have been published in the New Journal of Instrumentation [30] and the Nuclear Science Symposium and Medical Imagine Conference 2013 [11].